\documentclass[aps,twocolumn,showpacs,groupedaddress]{revtex4}
\usepackage{graphicx}

\begin{document}
\title{Dynamics of vortex matter in rotating two-species Bose-Einstein condensates}
\author{S. J. Woo, S. Choi, L. O. Baksmaty, and N. P. Bigelow}

\affiliation{Department of Physics and Astronomy, and Laboratory for Laser Energetics, University of Rochester, Rochester, NY 14627}

\date{\today}
\begin{abstract}
In a rotating two-component Bose-Einstein condensate (BEC), the traditional triangular vortex lattice can be replaced by a rectangular vortex lattice or even a structure characterized in terms of vortex sheets, depending on the interspecies interactions. 
We study the dynamics of this system by analyzing the Bogoliubov excitation spectrum.  
Excitations familiar to BEC vortex systems are found such as Tkachenko modes, hydrodynamic modes and surface waves, however, the complex two-component morphology also gives rise to new phenomena including shear flow between vortex sheets.
\end{abstract} 
\maketitle 

\newcommand{\figureA}
{
\begin{figure}
\centerline{\includegraphics[height=3.5cm]{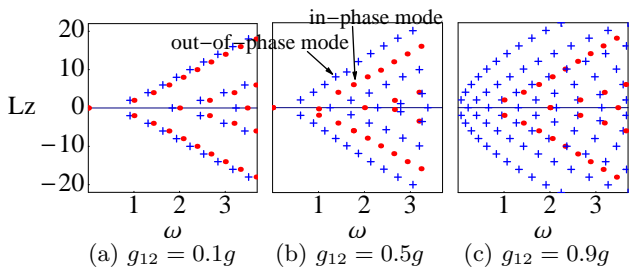}}\vspace{-.4cm}
\centerline{\hspace{.2cm} (a) $g_{12} = 0.1g$ \hspace{.3cm} (b) $g_{12} = 0.5g$ \hspace{.3cm} (c) $g_{12} = 0.9g$}
\caption{Red dots and blue crosses represent in-phase and out-of phase modes respectively.Out-of-phase modes gradually depart from in-phase modes decreasing energies as interspecies interaction increases.The spectrum of in-phase modes are not seriously affected by change of $g_{12}$.}
\label{PO_nonrotating_spectrum}
\end{figure}
}

\newcommand{\figureB}
{
\begin{figure}
\centerline{\includegraphics[width=8cm,height=6cm]{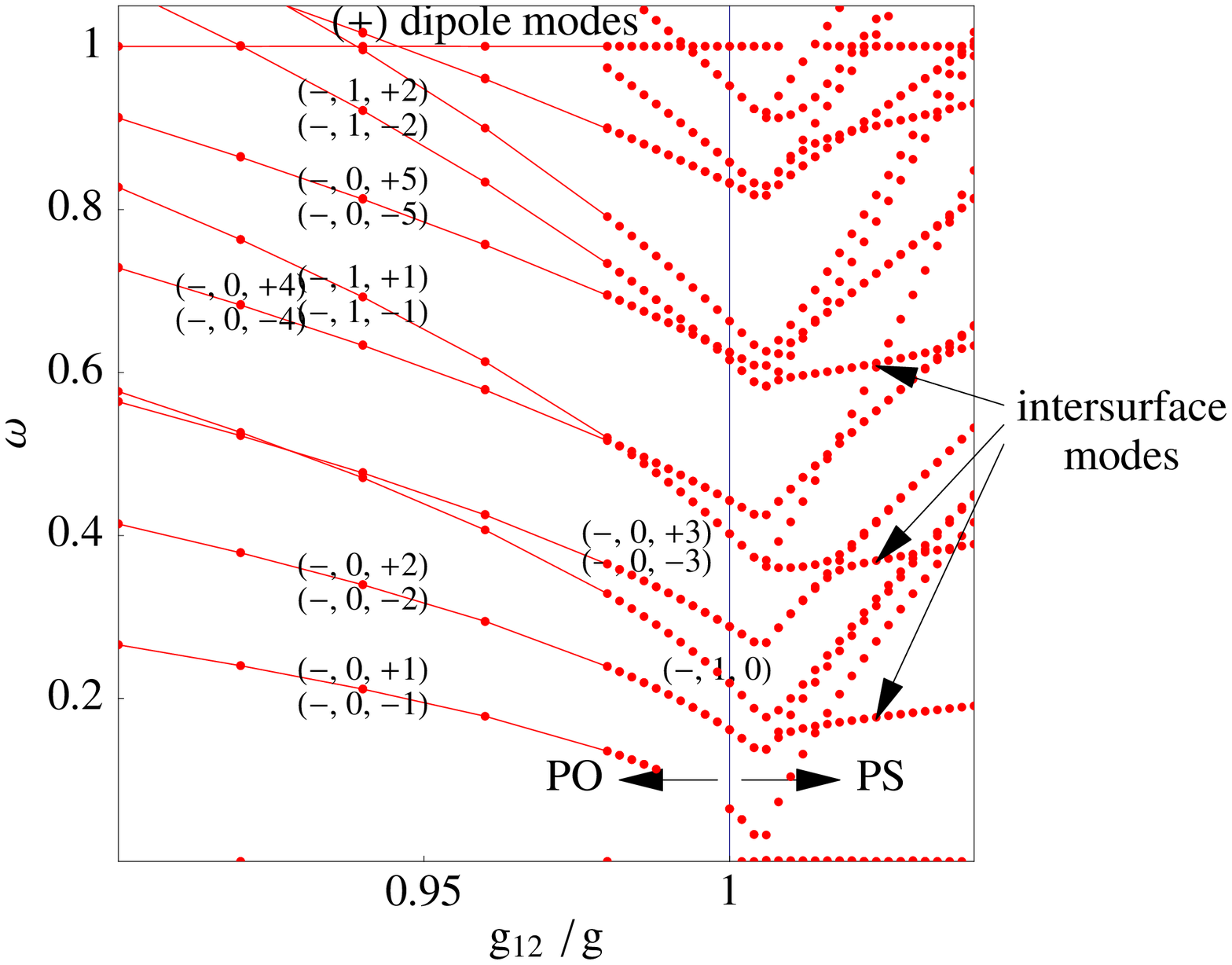}}
\caption{Energies of normal modes as a function of $g_{12}$ showing the transition from phase-overlapped to phase-separated state.The actual transition interaction strength $g^T_{12}$ is slightly larger than $g$ due to the kinetic energy contribution.}
\label{nonrotating_phasetransition}
\end{figure}
}

\newcommand{\figureC}
{
\begin{figure}
\centerline{\includegraphics[width=3cm]{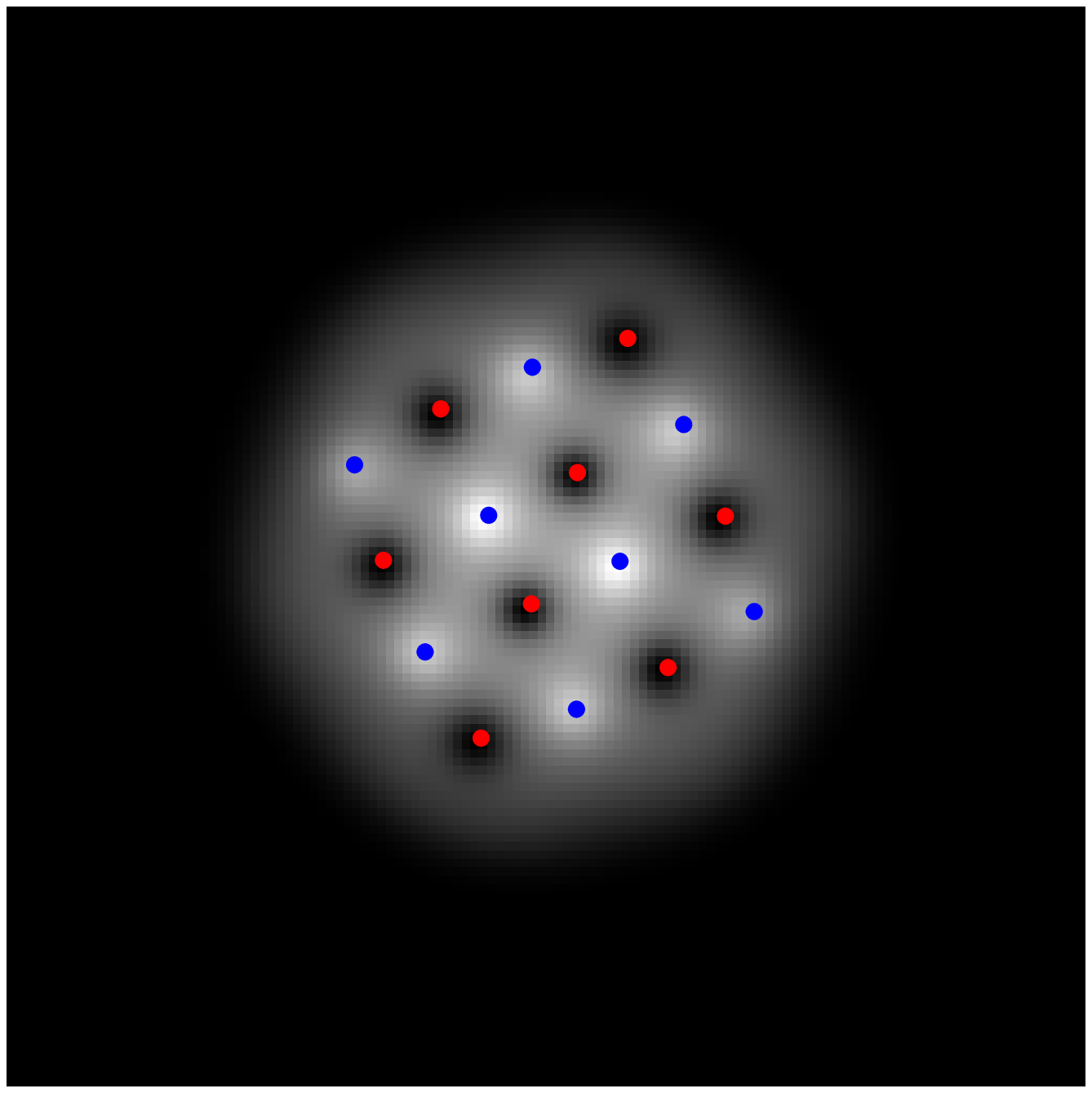} \hspace{-.5cm} \includegraphics[width=3cm]{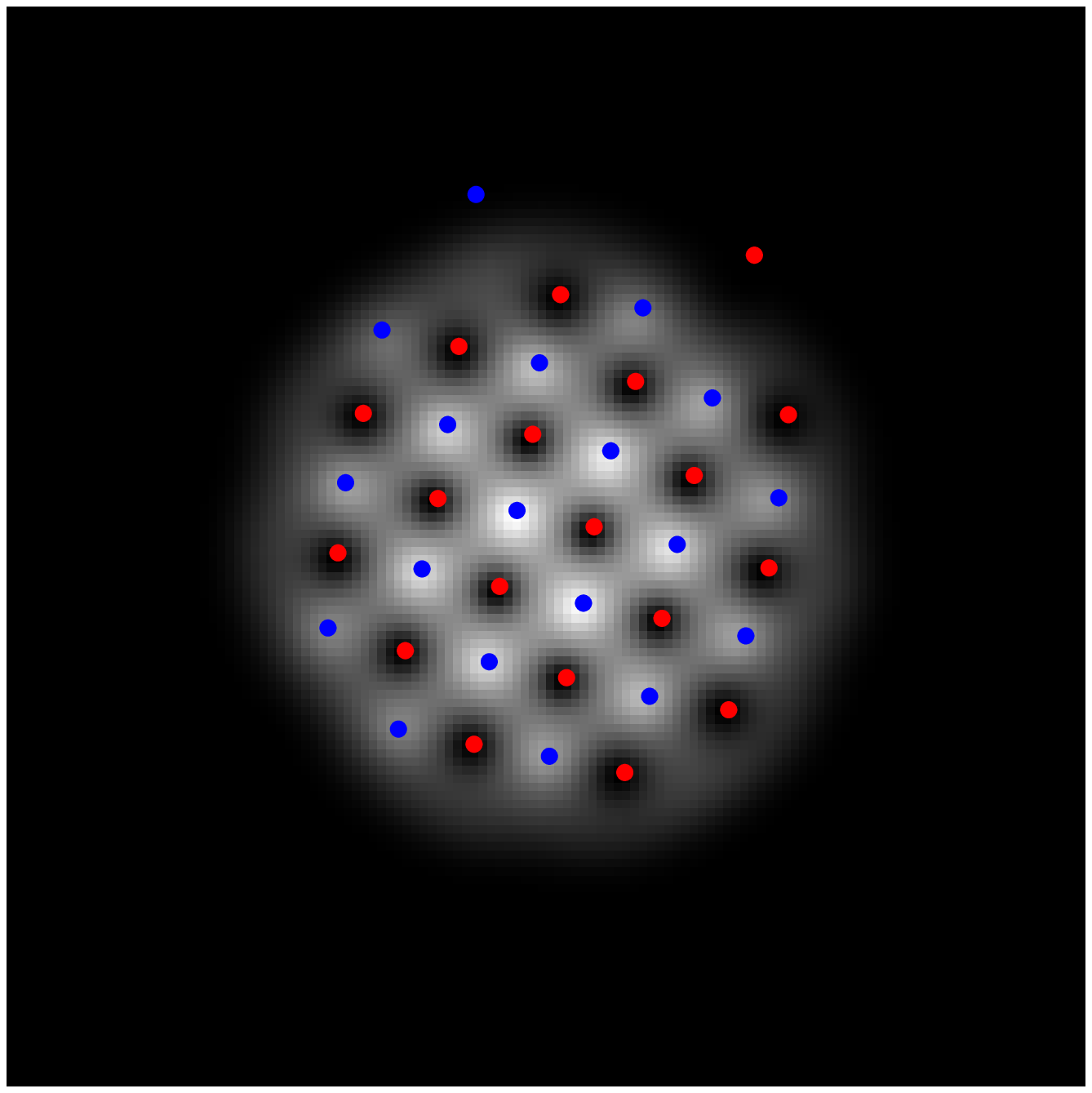} \hspace{-.5cm} \includegraphics[width=3cm]{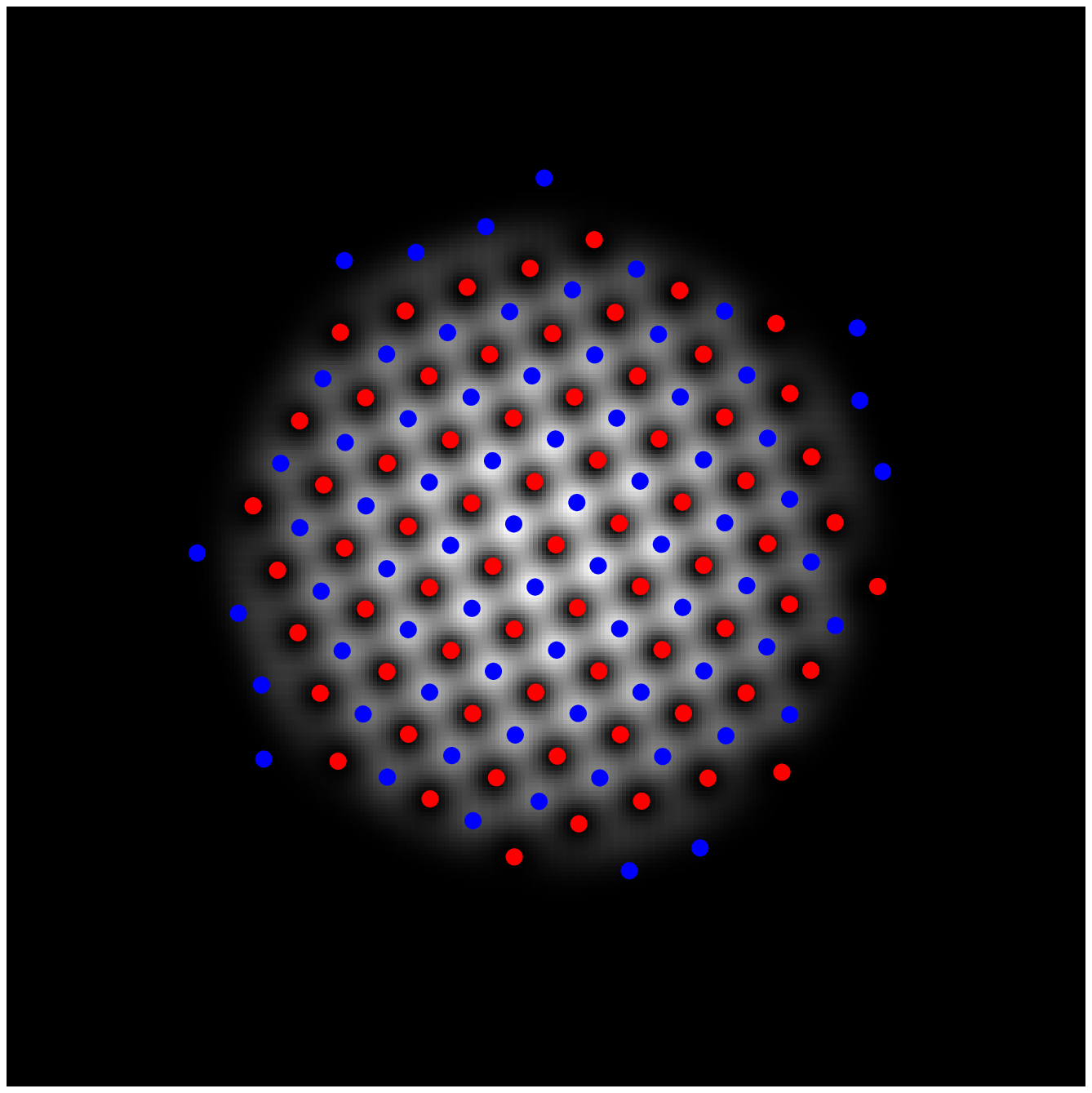}}
\centerline{(a) $\Omega = 0.4\omega_{\rm tr}$ \hspace{.5cm} (b) $\Omega = 0.6\omega_{\rm tr}$ \hspace{.5cm} (c) $\Omega = 0.9\omega_{\rm tr}$}
\caption{Three ground states depending on different rotational frequencies with $g_{12}=0.75g$.  Bright clouds represent one species with red dots the positions of vortices while blue dots are vortices of the other component.}
\label{rectangular gdstate}
\end{figure}
}

\newcommand{\figureD}
{
\begin{figure}
\centerline{\includegraphics[height=3.5cm]{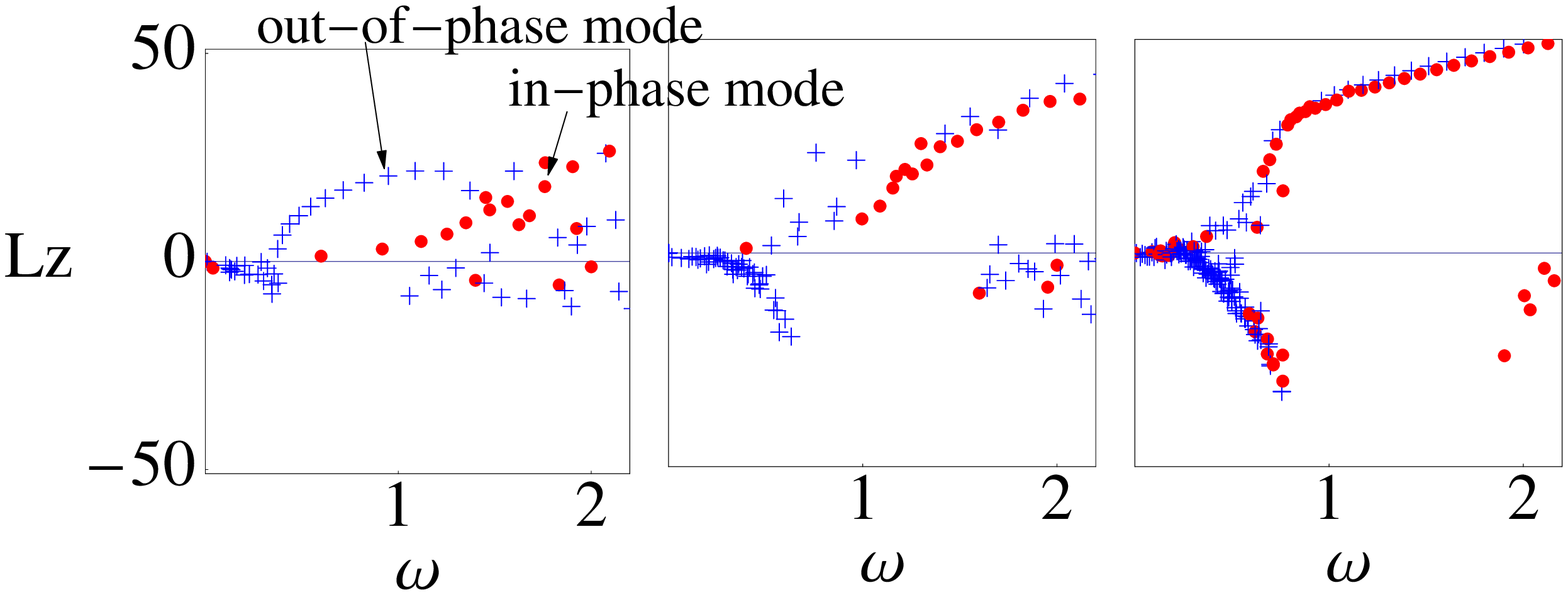}}\vspace{-.4cm}
\centerline{\hspace{.2cm} (a) $\Omega = 0.4\omega_{\rm tr}$ \hspace{.3cm} (b) $\Omega = 0.6\omega_{\rm tr}$ \hspace{.3cm} (c) $\Omega = 0.9\omega_{\rm tr}$}
\caption{Energy vs. angular momentum for the three ground states in Fig. \ref{rectangular gdstate}.  Each point represents a normal mode.Tkachenko modes are identified as negative angular momentum states in the low energy regime. Surface modes are divided into in-phase (red dots) and out-of-phase (blue crosses) modes.Note that the gap between spectral lines of in-phase and out-of-phase surface modes decreases at higher angular frequencies.}
\label{spectrum0.75}
\end{figure}
}

\newcommand{\figureE}
{
\begin{figure}
\centerline{\includegraphics[height=3.5cm]{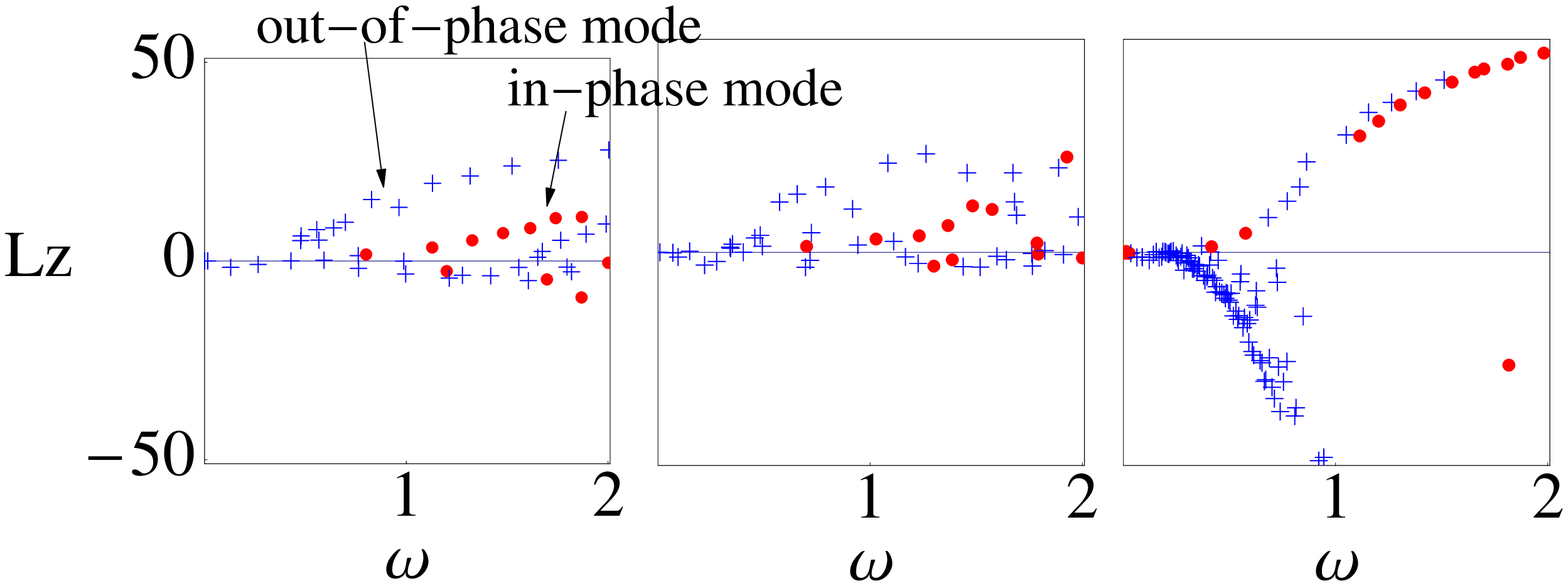}}\vspace{-.4cm}
\centerline{\hspace{.2cm} (a) $\Omega = 0.2\omega_{\rm tr}$ \hspace{.3cm} (b) $\Omega = 0.3\omega_{\rm tr}$ \hspace{.3cm} (c) $\Omega = 0.8\omega_{\rm tr}$}
\caption{Spectrum of three different vortex sheet states corresponding to Figs. [\ref{sheet gdstate}](a--c).}
\label{sheet_spectrum}
\end{figure}
}

\newcommand{\figureF}
{
\begin{figure}
\centerline{\includegraphics[width=3cm]{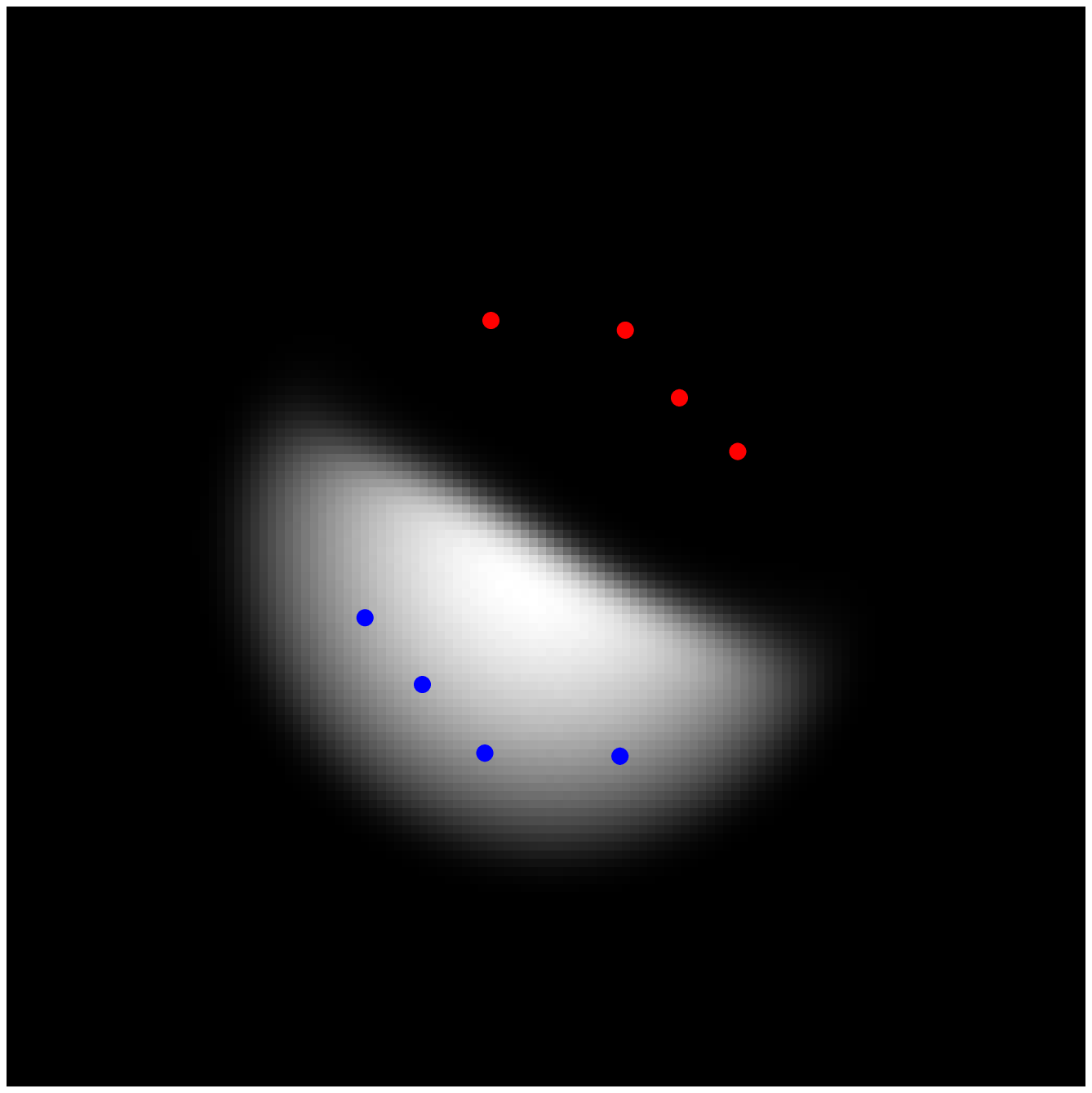} \hspace{-.5cm} \includegraphics[width=3cm]{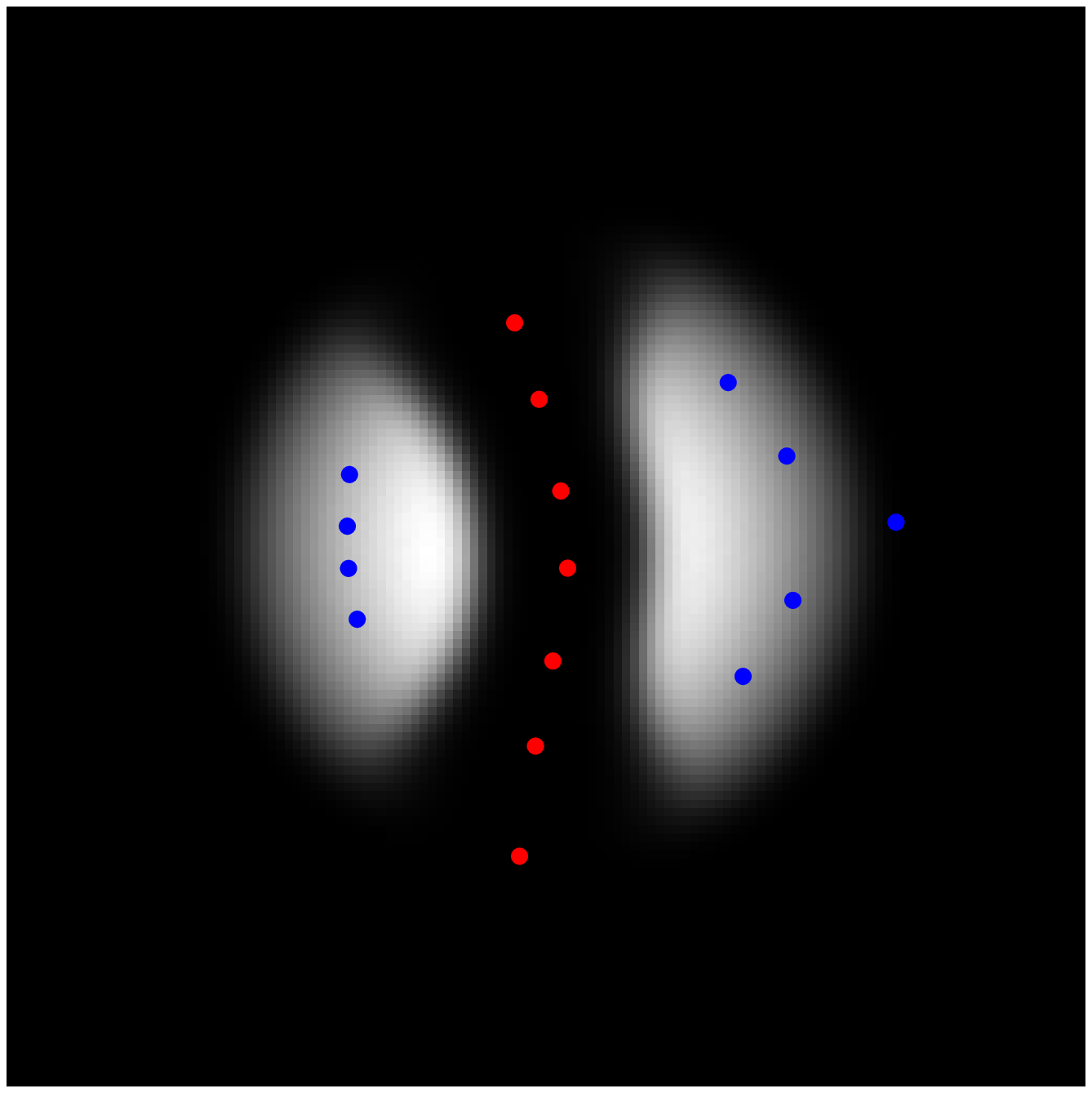} \hspace{-.5cm} \includegraphics[width=3cm]{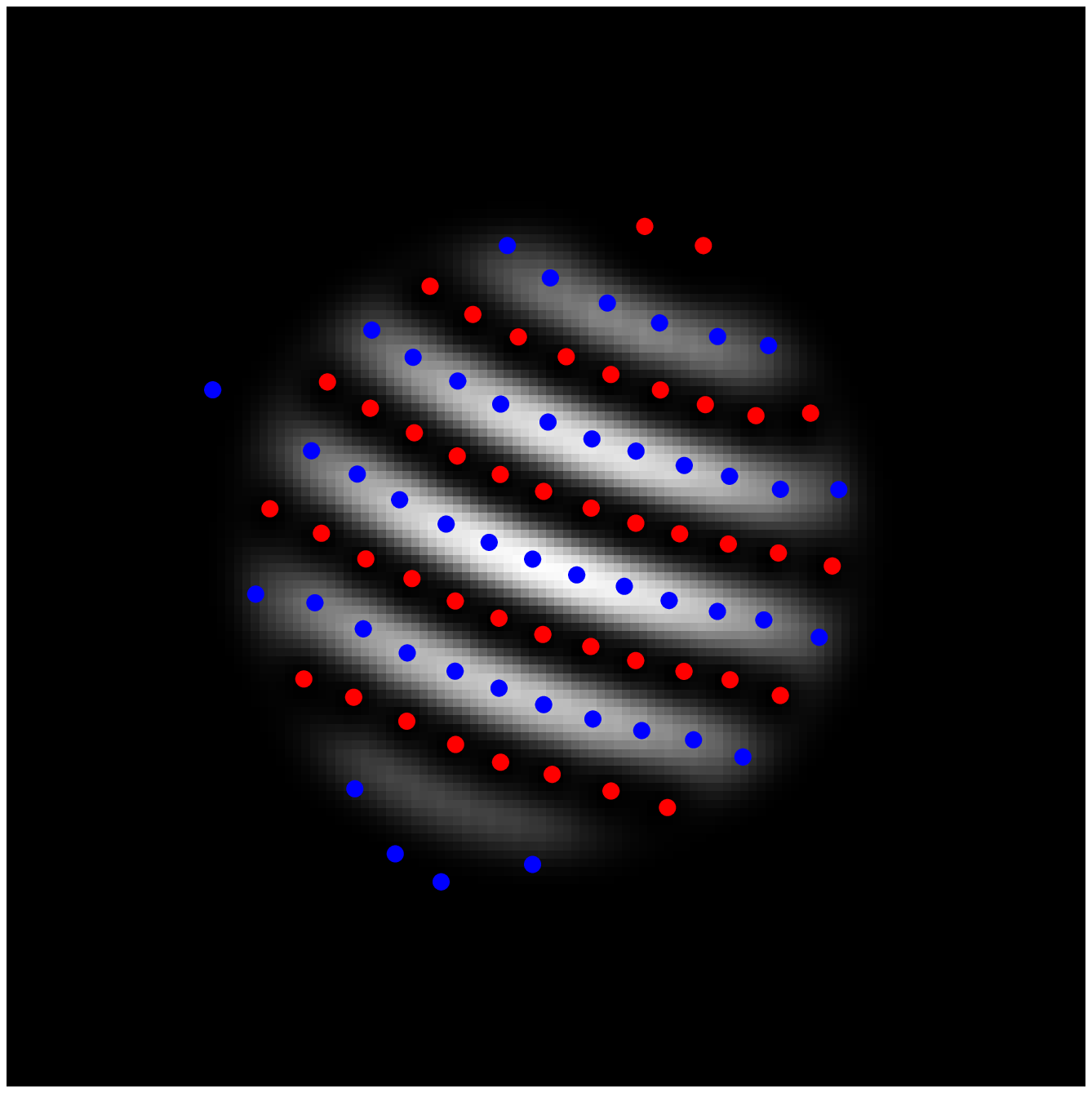}}\centerline{(a) $\Omega = 0.2\omega_{\rm tr}$ \hspace{.5cm} (b) $\Omega = 0.3\omega_{\rm tr}$ \hspace{.5cm} (c) $\Omega = 0.8\omega_{\rm tr}$}\caption{Three ground states depending on different rotational frequencies with $g_{12}=1.1g$.  Bright clouds represent one species with red dots the positions of vortices while blue dots are vortices of the other component.(a) is close to the nonrotating PS condensate except that it has phase singularities stabilized by the presence of the other component.(b) is the first emergence of a vortex sheet (red dots) that effectively divides the condensate into two pieces.(c) is a typical vortex sheet state for high rotational frequency.}
\label{sheet gdstate}
\end{figure}
}

\newcommand{\figureG}
{
\begin{figure}
\centerline{\includegraphics[width=3.5cm]{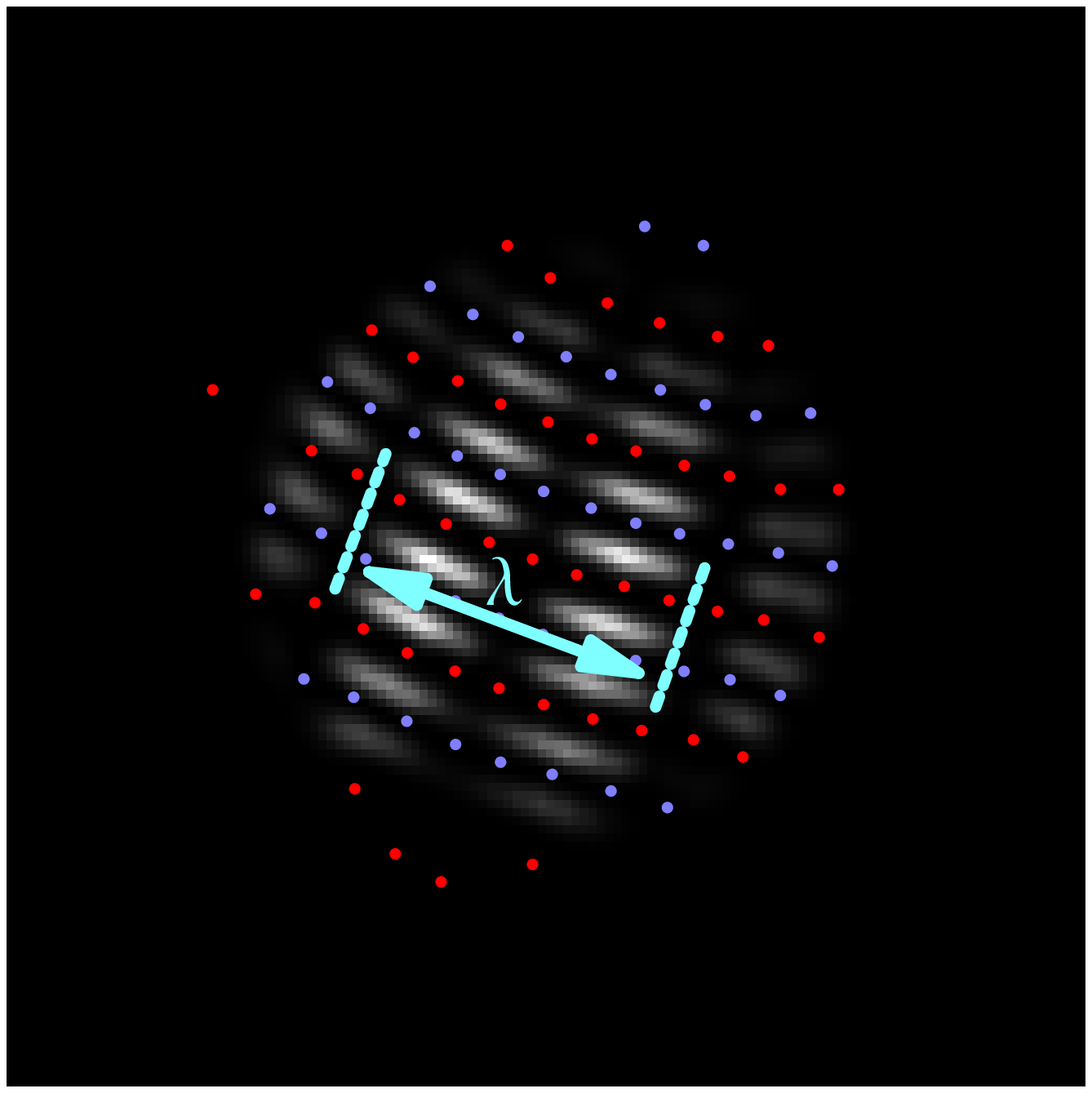}\hspace{.5cm}\includegraphics[width=3.5cm]{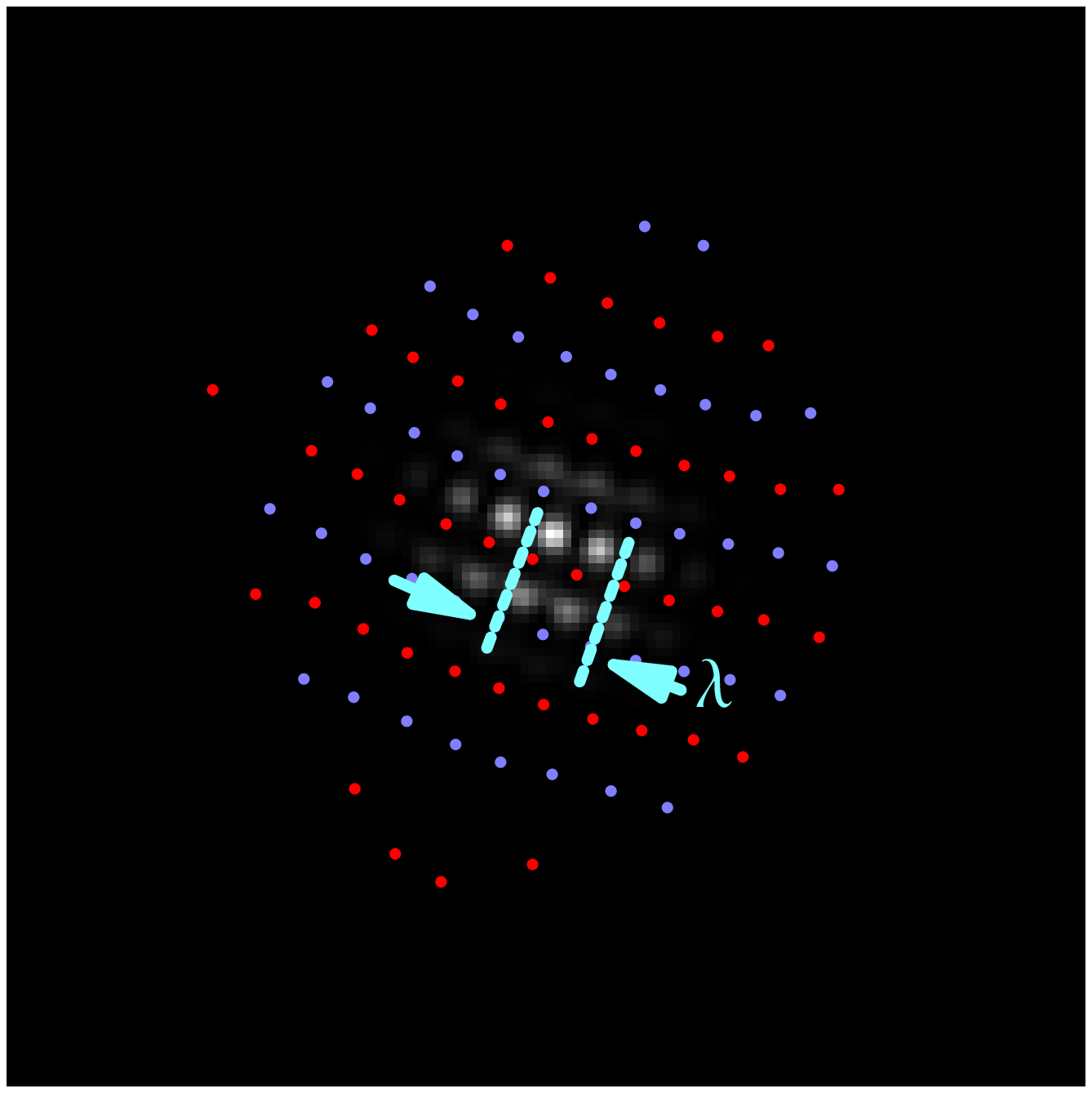}}\centerline{(a)\hspace{3.5cm}(b)}
\caption{Amplitude of density fluctuation $n'$ of Tkachenko modes with wave length (a) longer and (b) shorter than intersheet spacing.}
\label{anisotropic_Tk}
\end{figure}
}

\newcommand{\figureH}
{
\begin{figure}
\centerline{\includegraphics[width=5cm]{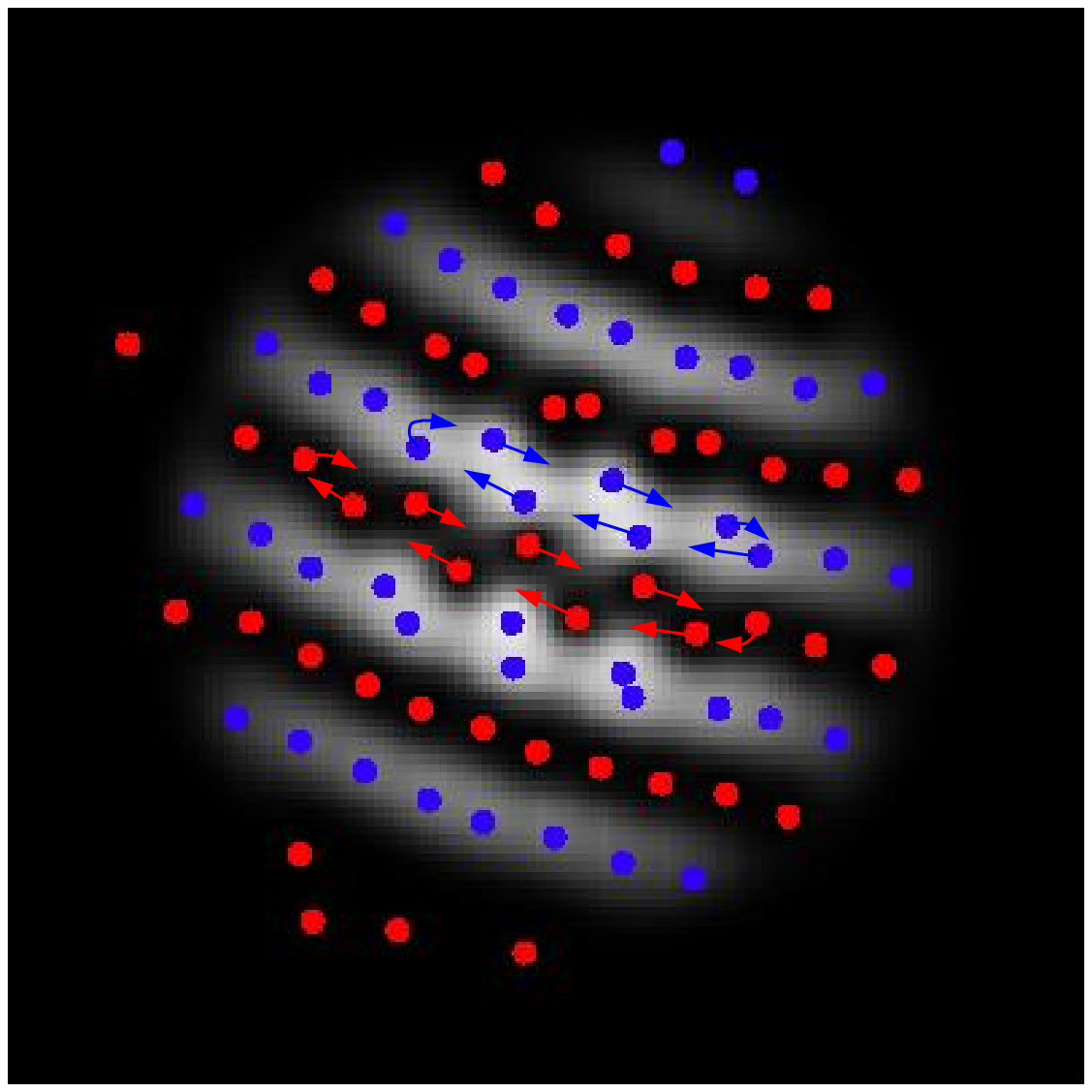}}
\caption{A snapshot of the most highly excited Tkachenko mode. Arrows are added to show the motion of vortices flowing along each sheet.}
\label{melting}
\end{figure}
}

\vspace{6mm}

Since the realization of vortices in trapped atomic Bose-Einstein condensate (BEC) \cite{Dalibard, Ketterle, JILA}, there have been extensive efforts to understand the dynamics of quantized vortices in a dilute bosonic system from both the theoretical and experimental view \cite{FetterReview,Dalibard_Hodby}. 
With the exciting developments in multi-species BEC, it is natural to ask about the nature of vortices in these systems.  
In the absence of rotation, it is known that the ground state of a two-species condensate goes through phase transition from a phase-overlapped (PO) to a phase-separated (PS) condensate depending on the interspecies interaction strengths \cite{Ao, Esry};the equivalent transition for a rotating system is that from a rectangular lattice to vortex sheets \cite{Kasamatsu}. 
The rectangular vortex lattice has been observed experimentally at JILA with rubidium hyperfine states as the two species \cite{Schweikhard}. 
The vortex-sheet configuration was also observed in a different context, $^3 \mbox{He-A}$, in which two different spin states of the Cooper pairs constitute the two species \cite{Parts}.
In this paper we discuss the excitations of the rotating two-species condensate in both the PO and PS regimes. 

With two mean field order parameters in a single system, the inter-species interaction energy, $g_{12}\int d{\bf r}\ \hat{\psi}^\dagger_2\hat{\psi}^\dagger_1\hat{\psi}_1\hat{\psi}_2$ is introduced, where $g_{12}$ is the interaction strength between species $1$ and $2$. 
We ignore any coherent coupling between species in this paper.
With this additional term in the free energy, the Gross-Pitaevskii equation (GPE) in a rotating frame \cite{Leggett} becomes ($i = 1,2$)\begin{equation}\label{2cGPE}\left({\cal H} + g_{i}|\psi_i|^{2} + g_{12}|\psi_{3-i}|^{2} - \Omega{\cal L}_z \right)\psi_i = \mu_i\psi_i,\end{equation}which can be solved to find the ground state of a condensate.
Here, ${\cal H}=-(\hbar^2/2M)\nabla^2+V_{\rm tr}({\bf r})$ and ${\cal L}_z=-i\hbar\partial/\partial\phi$.  
$\Omega$ is the rotational frequency of the reference frame and $g_i = 4\pi a_i\hbar^2/M_i$ is the interparticle interaction strength of species $i$  with the $s$-wave scattering length $a_i$. 
With this choice of parameters, in the absence of rotation, the transition point from a PO to PS condensate is found at $g_{12}=g$ \cite{Ao, Esry}.

The coupled Bogoliubov-de Gennes (BdG) equations \cite{Leggett} that describe linearized dynamics around the ground states $\psi_1$ and $\psi_2$ are ($i = 1,2$)\begin{eqnarray}\label{2cBdG}&&\left(\tilde{\cal H}_i-\Omega{\cal L}_z\right)u_{i} - g_i\psi_i^2v_{i} + g_{12}\psi_i n'_{3-i} = \hbar\omega u_{i},\\&&\left(\tilde{\cal H}_i+\Omega{\cal L}_z\right)v_{i} -g_i\psi_i^{*2}u_{i} - g_{12}\psi_i^* n'_{3-i} = -\hbar\omega v_{i}, \nonumber \end{eqnarray}where $\tilde{\cal H}_i={\cal H}-\mu_i+2g_i|\psi_i|^{2} + g_{12}|\psi_{3-i}|^{2}$ with $i=1,2$ and $n'_{i}=\psi_i^*u_{i} - \psi_iv_{i}$ is the density fluctuation for species $i$ \cite{DalfovoReview}.
We use a harmonic trap potential $V_{\rm tr}=M\omega_{\rm tr}^2(x^2+y^2)/2+M\omega_z^2z^2/2$.
We further assume $g_1=g_2\equiv g$ which is relevant for the experiment \cite{Schweikhard}.
The BdG equations are solved numerically by discretizing the eigenvalue problem, using finite element method with local cubic Hermite polynomials as the basis.
The solutions are analyzed by observing the dynamics of $\psi_i(t)=\psi_i^0 + u_i e^{-i\omega t} - v_i^* e^{i\omega t}$ \cite{Leggett} together with the density fluctuation $n_i'$. 
Here $\psi_i^0$ is the ground state solution of the $i$-th component. 
Our analysis assumes a quasi 2-dimensional geometry appropriate for a pancake shaped trapping potential with the dynamics along the $z$-axis frozen.

With $g_{12} < g$, the energy spectrum of a non-rotating PO condensate is separated into two branches with in-phase ($+$) and out-of-phase ($-$) modes. 
Each mode can be characterized using radial and angular quantum numbers, $n$ and $m$ just as is done for a single component condensate \cite{Fliesser, Ohberg, Woo}.
The overall density $n_t=\psi_1^*\psi_1 + \psi_2^*\psi_2$ for the out-of-phase modes does not oscillate significantly,which can be understood by noting that each species moves oppositely to the other, reminiscent of Landau's second sound in superfluid helium. 
To concretely distinguish the ($+$) modes from the ($-$) modes, the difference in the complex phase between two density fluctuations $n_1'$ and $n_2'$ can be quantified by defining $D \equiv \int d{\bf r} n_1^{'*} n_2' / \int d{\bf r}|n_1'||n_2'|$ which takes the value between $+1$ and $-1$:$D \sim 1$ for ($+$) modes and $-1$ for ($-$) modes.  
\figureA
Figure \ref{PO_nonrotating_spectrum} shows the departure of the ($-$) modes (blue crosses) from ($+$) modes (red dots) as $g_{12}$ increases from zero up to $g_{12}=g$. 
The frequencies of the ($+$) modes do not change significantly with the change of $g_{12}$. 
To interpret this, note that the ($+$) modes are barely affected by the interspecies interaction as the two species move together rather than against each other.
As in a single species case \cite{Woo} it is found that highly excited surface modes also exist that involve a circle of phase singularities flowing together with a density wave around the edge of the condensate.
For the ($+$) modes the phase singularities for each of the two species overlap one another,  while for the ($-$) modes, they alternate around the edge of the condensate.  
For a given quantum number $m$, the ($-$) mode has lower energy compared to the ($+$) mode since, for $g_{12} > 0$, the separation between phase singularities from different species is energetically preferred. 

\figureC
When a PO BEC is rotated, it develops two interlacing rectangular vortex lattices -- one condensate fills up the density holes made by the vortices in the other component [Fig. \ref{rectangular gdstate}]. 
It is interesting to note that, because of the finite number of vortices corresponding to a given rotational frequency, the azimuthal symmetry of the vortex lattice structure differs for different rotational frequencies. 
As in a rotating single component BEC, two types of normal modes were found to emerge once the system contains vortices: the excitations of the vortex structure (Tkachenko modes) and fluid density fluctuations of the underlying condensate (hydrodynamic modes) including surface modes. 
\figureD
The surface modes constitute one of the main types of excitations in the lower energy regime as shown in Fig. \ref{spectrum0.75}, the energy vs. angular momentum plots for $g_{12}=0.75g$.   
For the Tkachenko modes, one might expect that, with two species present, there would be two branches of normal modes, the ``optical'' and ``acoustic'' modes similarly to a solid state crystal lattice with two constituents \cite{Kittel, Oktel}. 
Numerical calculation, however, indicates that the distinction between these two branches is not evident. 
This is due to the fact that in the BEC system the interaction between vortices in different species is caused by density-density interactions and hence is far smaller than the inter-vortex interaction within the same species dominated by the kinetic energy term.

Figure \ref{spectrum0.75} shows that the discrepancy between in-phase (red dots) and out-of-phase (blue crosses) modes becomes negligible as $\Omega$ increases.
This is because the phase singularities involved in the surface modes no longer have either clean overlap or separation due to the presence of vortices near the edge of the condensate.  
When the vortex lattice structure near the edge of the condensate is azimuthally asymmetric, the surface wave of one species interweaves the surface wave of the other species.
For the case of azimuthally symmetric vortex lattice structure, however, the phase singularities that constitute the surface wave of one species is uniformly closer to the edge of the condensate than those of the other species due to the repulsive interaction between the outer vortices of the vortex lattice and the phase singularities (vortices) in the surface wave. 
It was found that in such cases, the ratio of the quasiparticle populations from the two species integrated through the whole condensate departs from unity dramatically resulting in only one of the species fluctuating while the other component remains nearly stationary: the quasiparticle population of the species whose excitation is closer to the edge of the condensate dominates the other. 
Such quasiparticle population imbalance of the surface modes are a new feature of the rotating condensate, and can provide a potential diagnostic tool for the vortex lattice structure.  

\figureB
For $g_{12} > g$, due to the broken rotational symmetry of the non-rotating PS BEC ground state, the degeneracies of $+m$ and $-m$ states of the PO ($g_{12} < g$) state for the out-of-phase modes no longer persists.
It has been suggested that due to the surface tension there might therefore be a new branch of normal modes on the surface that separates the two species \cite{Ao}.
Our numerical solution shows that, for a strongly phase-separated condensate, such modes exist and indeed dominate the lower energy regime.
These modes are not supplemental to the hydrodynamic excitations; instead they can be extrapolated from the out-of-phase hydrodynamic modes of PO BEC by changing $g_{12}$ continuously. 
Figure \ref{nonrotating_phasetransition} shows the change in spectrum as the interspecies interaction goes through the transition point.
The actual transition point is slightly larger than $g$ due to the kinetic energy contribution.
Note that, for $g_{12} > g$, there are different slopes for different branches, as in a single-species BEC with broken rotational symmetry \cite{Woo2}.
The group of lines with the lowest slope ($\sim 0$) corresponds to the intersurface wave propagating between species.  
They have different numbers of nodes depending on the intercept of the spectral lines with energy axis.
The frequencies of these modes do not noticeably depend on any form of surface tension (or $g_{12}$) as they do not penetrate far into the condensate.
The modes with non-zero slopes i.e. those with energies which depend on $g_{12}$  have additional nodes in the density fluctuation along the axis perpendicular to the inter-surface and the slope is determined by the number of such nodes. 

When the PS condensate is rotated, the formation of vortex sheets in the system effectively slices the condensate into several layers. 
By forming vortex sheets, the condensate achieves maximal phase separation compared to a lattice. 
It is notable that, for a given vortex density (which depends on the mass of the atom and the rotational speed of the system) one can achieve much smaller intervortex spacing within a vortex sheet when compared to a lattice geometry. 
We find that, for the case of vortex sheets, in contrast to the unique structure of a vortex lattice, there exist different metastable vortex sheet configurations with almost the same energies.  
We understand this by noting that the energy of the system is mainly determined by two length scales, the intervortex spacing ($d_v$) within a vortex sheet and the  intersheet spacing ($d_s$), rather than the shape of the vortex sheets \cite{Kasamatsu}.
\figureF
However, since any bending or defects in the sheet configuration still has some energy cost, the state with straight vortex sheets corresponds to the lowest energy for our parameters and has been used for the analysis [Fig. \ref{sheet gdstate}].

\figureE
Figure \ref{sheet_spectrum} shows the frequency vs. angular momentum curve for $g_{12} = 1.1g$.
The Tkachenko modes are clearly identified in the lower energy regime especially for the fast rotating case (c), while hydrodynamic excitations are found throughout the whole region of energy.
It is obvious that the wavelength of the Tkachenko mode should be bounded below by $d_v$. 
We find that, different from the triangular or rectangular vortex lattices, Tkachenko modes are anisotropic due to the inhomogeneity of the vortex matter. 
The resulting density fluctuations are now concentrated along the vortex sheet. 
Figure \ref{anisotropic_Tk} shows $|n^{\prime}|^2$ for two Tkachenko modes, with wavelengths longer and shorter than the $d_s$ respectively.
The long-wavelength Tkachenko mode [Fig. \ref{anisotropic_Tk}(a)] is excited homogeneously throughout the whole region of the vortex matter since the wavelength is longer than $d_s$, while the short-wavelength Tkachenko mode [Fig. \ref{anisotropic_Tk}(b)] is excited mainly along a single sheet with the most anisotropy since waves with wavelength shorter than $d_s$ cannot travel across from one sheet to another.

\figureG
\figureH
We have also observed the breakdown of stable vortex positions associated with the excitations of certain Tkachenko modes. 
This happens when the typical vortex precession radius becomes the same order as $d_v$ which, in turn, is comparable to the healing length of a vortex within a vortex sheet.
For some highly excited Tkachenko modes, the relative precession motion of the nearest neighbor vortices are out of phase such that the vortices ``collide" with one another when the precession radius is larger than $d_v/2$.  
The result is an exchange of vortex positions in the lattice site. 
The successive exchange of vortex positions leads to a shear flow of ``delocalized'' vortices along the vortex sheet [Fig. \ref{melting}]. 
We have confirmed that this flow can happen stably and without experiencing any significant nonlinear decay within a time-dependent GPE simulation.
This vortex flow gives rise to ``inner" surface waves along the edge of the layers of the condensate and therefore such modes may provide a connection between surface modes and Tkachenko modes.
In general these ``inner" surface waves cannot flow out naturally to the outer surface of the overall BEC since the frequencies and wavelengths of the ``inner" and ``outer" surface waves do not necessarily match.
However, for some modes, we have observed that the two types of surface waves can be connected through the generation of temporary vortex-antivortex pair which provides the necessary frequency matching.

In summary, we have studied the normal mode dynamics of a rotating two-species BEC and found new features that result from the different ground state structures. 
For PO state with rectangular lattices, we found that quasiparticle population imbalance may provides a diagnostic tool for the vortex lattice structure.
For PS state with vortex sheets, we observed the breakdown of ground state vortex positions through shear flow of vortices within the linearized regime.
We note that as compared to a vortex lattice with similar vortex density, the shorter intervortex distance in the sheets allows observation of vortex collisions at lower rotation frequency.
The potential link between vortex delocalization and vortex lattice melting will be a topic of future investigation.

This work is supported by the NSF, DOE, ONR and ARO.

\end{document}